\documentclass[a4paper,twocolumn,showpacs,pra,superscriptaddress,eqsecnum]{revtex4} 
\usepackage{graphicx}
\usepackage{dcolumn} 
\usepackage{bm} 
\usepackage{amsmath}
\usepackage{amsfonts} 
\usepackage{amssymb}
\usepackage{color}

\usepackage[normalem]{ulem}

\begin{document}
\title{Thermodynamics of a trapped unitary Fermi gas}

\author{R. Haussmann}
\affiliation{Fachbereich Physik, Universit\"at Konstanz, D-78457 Konstanz, Germany}

\author{W. Zwerger}
\affiliation{Department of Physics, MIT-Harvard Center for Ultracold Atoms and
Research Laboratory of Electronics, Massachusetts Institute of Technology, Cambridge,
Massachsusetts, 02139, USA}
\affiliation{Technische Universit\"at M\"unchen, James-Franck-Strasse, D-85748 Garching, Germany}

\date{\today}

\begin{abstract}
Thermodynamic properties of an ultracold Fermi gas in a harmonic trap 
are calculated within a local density approximation, 
using a conserving many-body formalism for the BCS to BEC crossover problem, 
which has been developed by Haussmann \textit{et al.}\ 
[Phys.\ Rev.\ A \textbf{75}, 023610 (2007)]. We focus on 
the unitary regime near a Feshbach resonance and determine the local density 
and entropy profiles and the global entropy $S(E)$ as a function of the total 
energy $E$. Our results are in good agreement with both experimental data 
and previous analytical and numerical results for the thermodynamics of the unitary 
Fermi gas. The value of the Bertsch parameter at $T=0$ and the superfluid 
transition temperature, however, differ appreciably. We show that, well 
in the superfluid regime, removal of atoms near the cloud edge enables cooling 
far below temperatures that have been reached so far.

\end{abstract}

\pacs{03.75.Ss, 03.75.Hh, 74.20.Fg}

\maketitle

\section{Introduction}
\label{section_1}
The BCS to BEC crossover problem of a Fermi gas  with an adjustable 
attractive interaction has  been investigated theoretically for quite some time \cite{Eagles,Leggett,NSR,Drechsler,Randeria1,Haussmann1,Haussmann2}. 
For low temperatures the gas is superfluid and, in the case of s-wave 
interactions, it exhibits a smooth crossover from the well known
BCS regime of weakly bound Cooper pairs to the BEC regime of tightly bound 
bosonic dimers with a residual repulsive interaction \cite{Drechsler,Randeria1,Petrov1}. 
In recent years, this crossover has been realized experimentally using ultracold Fermi 
gases in optical traps, where the interaction can be tuned using Feshbach resonances
\cite{OHara,Greiner03,Jochim03,Zwierlein03,Regal04,Bartenstein04,Zwierlein04,
Kinast04,Bourdel04}.
In the experimentally relevant case of so called broad Feshbach resonances, 
which is in principle always realized in the dilute limit,
the physical properties of the homogeneous gas at equal densities for both
spin components are described by only two dimensionless parameters:
the  interaction strength $v=1/k_Fa$ and the temperature 
$\theta=k_B T/\varepsilon_F$. Here, $a$ is the $s$-wave scattering length which 
fully characterizes interactions in the dilute, ultracold limit, while the scales 
for length and energy are determined by the Fermi wave number $k_F=(3\pi^2 n)^{1/3}$ 
and the Fermi energy $\varepsilon_F=\hbar^2 k_F^2 / 2m$, respectively, where $n=N/V$ 
is the particle density. 

A particularly interesting regime is located near the Feshbach resonance, 
where the scattering length $a$ is infinite. At this point and, more generally, 
in the so-called \textit{unitary regime} where $k_F\vert a\vert \gg 1$, 
the dimensionless interaction parameter $v$ disappears from the problem. All 
thermodynamic quantities are therefore universal functions of the dimensionless 
temperature $\theta=k_B T/\varepsilon_F$ \cite{Ho04}. On a microscopic level, 
the unitary gas exhibits a particular kind of scale invariance, similar to a gas 
with purely inverse square two-particle interactions \cite{Werner06}. More 
generally, as shown by Nikolic and Sachdev \cite{Nikolic07}, universality is not 
restricted to the unitary regime. It is tied to the fact that the 
unitary balanced gas at zero density is an unstable fixpoint with only three 
relevant perturbations. Since there is no small expansion parameter, the unitary 
regime is the most challenging one from a theoretical point of view. In addition, 
it is in fact precisely this regime which is accessible experimentally (see 
e.g.\ the recent review articles by Ketterle and Zwierlein \cite{KZ08}, by 
Bloch \textit{et al.}\ \cite{BDZ08} and by Giorgini \textit{et al.}\ \cite{GPS08}).

In a recent paper \cite{Haussmann3}, we have presented a field theoretic approach 
for the thermodynamics of the BCS to BEC crossover, which is based on the 
formalism developed by Luttinger-Ward \cite{LW60} and DeDominicis-Martin 
\cite{DM64}. In the following, this approach for the homogeneous gas is 
applied to calculate the thermodynamic properties of the trapped Fermi gas,
using a local density approximation. We compare our results 
with a recent experiment by Luo \textit{et al.}\ \cite{Luo07} and with recent 
theories \cite{HLD06,HDL07,Bulgac07}. In particular, we provide 
results for the entropy as a function of temperature, 
which allows to do reliable thermometry for the trapped unitary gas 
and also gives a precise value for the critical temperature and the 
associated entropy per particle. In addition, we show that starting well 
in the superfluid regime, much lower temperatures and entropies can be reached 
by removing atoms from the edge of the cloud, which carry most of the entropy.

\section{Local density approximation}
\label{section_2}
In our previous paper we have calculated the thermodynamic quantities for the 
homogeneous system. At a given particle density $n=N/V$, these are the internal 
energy per particle $u=U/N$ and the entropy per particle $s=S/N$. 
The Fermi wave number $k_F=(3\pi^2 n)^{1/3}$ and the Fermi  energy 
$\varepsilon_F=\hbar^2 k_F^2 / 2m$ can be used as scale factors in order to 
make the thermodynamic quantities dimensionless. 

In the presence of an external confining potential 
$V(\mathbf{r})$, the particle density $n(\mathbf{r})$ is non-uniform.
Within a local density approximation (LDA), thermodynamic quantities 
like pressure $p(\mathbf{r})$, chemical potential $\mu(\mathbf{r})$,
entropy per particle $s(\mathbf{r})$ or the internal energy per particle 
$u(\mathbf{r})$ are then also spatially varying, being determined by the 
corresponding equilibrium values in the uniform system
evaluated at the local density $n(\mathbf{r})$. The local density approximation 
neglects the dependence of thermodynamic properties on density gradients.
At zero temperature, it is essentially a zeroth order semiclassical approximation 
\cite{Brack_1}. It is valid as long as the local Fermi wave number $k_F(\mathbf{r})$ 
times the oscillator length $\ell_0=(\hbar/m\omega)^{1/2}$ defined by the 
characteristic frequency $\omega$ of the confining potential is much larger 
than one. Except near the edge of the cloud, where the density approaches zero, 
this condition is well justified for most of the experiments because typical 
Fermi energies $\varepsilon_F$ are of the order of several kHz, while the
trapping frequencies are around $\omega\approx 100\,\mathrm{Hz}$ or smaller.
Specifically, near the trap center $k_F(\mathbf{0})\ell_0\simeq N^{1/6}$, 
where $N$ is the total particle number in a trap.  As will be shown below, the 
finite size corrections to the ground state energy are of relative order
$(3N)^{-2/3}$ in a harmonic trap. They are therefore negligible, at least for 
global observables, for the typical particle numbers in experiment, where 
$N \approx 1.3 \times 10^5$ \cite{Luo07}. This conclusion is also
supported by a recent comparison of LDA with a numerical solution of the 
Bogoliubov-DeGennes equations \cite{LHD07}.

Within the LDA, the global thermodynamic quantities of the trapped Fermi 
gas are obtained by integrating over the whole space. Specifically, we define
\begin{eqnarray}
N &=& \int d^3r\ n(\mathbf{r}) \ ,
\label{Eq:particle_number} \\
E_\mathrm{pot} &=& \int d^3r\ n(\mathbf{r}) \ V(\mathbf{r}) \ ,
\label{Eq:potential_energy} \\
U &=& \int d^3r\ n(\mathbf{r})\ u(\mathbf{r}) \ ,
\label{Eq:internal_energy} \\
E &=& U + E_\mathrm{pot} = \int d^3r\ n(\mathbf{r})
\ [u(\mathbf{r})+V(\mathbf{r})] \ , \qquad 
\label{Eq:total_energy} \\
S &=& \int d^3r\ n(\mathbf{r})\ s(\mathbf{r}) \ ,
\label{Eq:entropy}
\end{eqnarray}
as the particle number, the potential energy, the internal energy, the total
energy, and the total entropy. Here the particle density acts like a 
distribution function to define averages over the trap.

In an optical trap where the laser intensity profiles are Gaussian functions,
the confining potential $V(\mathbf{r})$ is given by an anisotropic Gaussian function. 
Close to the center of the trap this potential can be approximated by an anisotropic 
harmonic function
\begin{equation}
V(\mathbf{r}) = \frac{1}{2}m(\omega_x^2 x^2 + \omega_y^2 y^2 + \omega_z^2 z^2)
\label{Eq:harmonic_potential}
\end{equation}
where $m$ is the mass of the atoms and $\omega_x$, $\omega_y$, $\omega_z$ 
are the harmonic oscillator frequencies in the three spatial directions. In the
following, we use the harmonic potential \eqref{Eq:harmonic_potential} and neglect the
anharmonic terms. Since in LDA $\mu(\mathbf{r})$ is the chemical potential relative 
to the potential $V(\mathbf{r})$, in thermal equilibrium the total chemical potential 
$\mu_\mathrm{tot} = \mu(\mathbf{r}) + V(\mathbf{r})$ is constant which implies the 
condition 
\begin{equation}
\mu(\mathbf{r}) + V(\mathbf{r}) = \mu(\mathbf{0}) \ .
\label{Eq:chempot_condition}
\end{equation}
This equation together with the requirement of a constant temperature $T$ determines 
the spatial dependence of all local thermodynamic quantities. The particle density 
$n(\mathbf{r})$ implies a local Fermi wave number $k_F(\mathbf{r})$ and a local 
Fermi energy $\varepsilon_F(\mathbf{r})$. As a consequence the dimensionless 
parameters $v(\mathbf{r})=1/k_F(\mathbf{r})a$ and 
$\theta(\mathbf{r})=k_B T/\varepsilon_F(\mathbf{r})$ depend on the local 
position in the trap.

It is convenient to define the weighted radial coordinate $r$ by
\begin{equation}
\omega^2 r^2 = \omega_x^2 x^2 + \omega_y^2 y^2 + \omega_z^2 z^2
\label{Eq:weighted_radial_coordinate}
\end{equation}
where $\omega=(\omega_x \omega_y \omega_z)^{1/3}$ is the average harmonic frequency. 
With this definition the confining potential acquires the simple form 
$V(\mathbf{r})=V(r)=\frac{1}{2}m\omega^2r^2$. As a consequence, the anisotropy of
the trap becomes irrelevant because all local quantities $n(\mathbf{r})=n(r)$ etc.\ 
only depend on the weighted radial coordinate $r$. Rewriting the space integrals
in Eqs.\ \eqref{Eq:particle_number}-\eqref{Eq:entropy} in terms
of $r$, this applies also to all thermodynamic quantities derived from the local
density. A convenient measure for the overall length scale in a trap is provided by
the Thomas-Fermi radius $R_{TF}=(24N)^{1/6}(\hbar/m\omega)^{1/2}$ of the confined 
{\it non-interacting} Fermi gas at zero temperature with a given total particle number
$N$. Similarly, as a characteristic scale for the energy we define the corresponding 
Fermi energy  $E_F=(3N)^{1/3}\hbar\omega$ of the non-interacting gas.

\section{Unitary regime at zero temperature}
\label{section_3}
The thermodynamic quantities can be made dimensionless by considering the ratios
$\mu(\mathbf{r})/\varepsilon_F(\mathbf{r})$, $u(\mathbf{r})/\varepsilon_F(\mathbf{r})$, 
and $s(\mathbf{r})/k_B$. These ratios depend on the space coordinate $\mathbf{r}$ only 
implicitly via the dimensionless parameters $v(\mathbf{r})=1/k_F(\mathbf{r})a$ and 
$\theta(\mathbf{r})=k_B T/\varepsilon_F(\mathbf{r})$. A particular case is the unitary  
gas at zero temperature, where both parameters vanish identically $v(\mathbf{r})=0$, 
$\theta(\mathbf{r})=0$. Using standard thermodynamic relations, it is straightforward to 
show  that all thermodynamic quantities can be expressed in terms of a {\it single} 
dimensionless parameter, the so-called Bertsch parameter $\xi$ \cite{Bertsch_1,Baker_1} 
in the form 
\begin{equation}
\frac{\mu(\mathbf{r})}{\varepsilon_F(\mathbf{r})}=\xi \ , \qquad
\frac{u(\mathbf{r})}{\varepsilon_F(\mathbf{r})}=\frac{3}{5}\xi \ . \qquad
\label{Eq:dimensionless_ratios}
\end{equation}
These relations hold both for an ideal Fermi gas, where $\xi=1$ and also at
the unitarity point, where $\xi$ has a  nontrivial value \cite{Ho04,Nikolic07}. 
In our previous work \cite{Haussmann3}, 
we have calculated both ratios independently and obtained 
$\mu(\mathbf{r})/\varepsilon_F(\mathbf{r})=0.358$ and
$u(\mathbf{r})/\varepsilon_F(\mathbf{r})=0.210$. The first ratio implies a Bertsch
parameter $\xi=0.358$, while the second ratio implies $\xi=0.351$. These two values
differ by about $2.0\%$. Consequently, the relation 
$u(\mathbf{r})/\mu(\mathbf{r})=3/5$, which is valid both for an ideal and a unitary 
Fermi gas, is satisfied up to an error of $2.0\%$.

Inserting the harmonic potential \eqref{Eq:harmonic_potential} into Eq.\ 
\eqref{Eq:chempot_condition} and using the weighted radial coordinate 
\eqref{Eq:weighted_radial_coordinate}, we obtain the chemical potential
\begin{equation}
\mu(\mathbf{r}) = \mu(\mathbf{0}) \ [ 1 - r^2 / r_{TF}^2 ]
\label{Eq:chemical_potential_2}
\end{equation}
where $r_{TF}$ is the Thomas Fermi radius of the unitary gas at 
zero temperature. The dimensionless ratios \eqref{Eq:dimensionless_ratios} imply 
similar functional forms for the other quantities, e.g.\ 
$k_F(\mathbf{r}) =k_F(\mathbf{0}) \ [ 1 - r^2 / r_{TF}^2 ]^{1/2}$ for the local Fermi 
wavevector. These expressions are valid only for $r<r_{TF}$, because the particle 
density $n(\mathbf{r})$ is nonzero only in this case and zero otherwise. Evidently, 
the Thomas-Fermi radius $r_{TF}$ of the unitary Fermi gas is the only parameter which 
determines the spatial dependence of the thermodynamic quantities. It is related to 
the Thomas-Fermi radius $R_{TF}$ of the ideal Fermi gas by the Bertsch parameter via
\begin{equation}
r_{TF}/R_{TF} = [\mu(\mathbf{0})/\varepsilon_F(\mathbf{0})]^{1/4} = \xi^{1/4} \ .
\label{Eq:Thomas_Fermi_radius}
\end{equation}
Using the dimensionless ratio $\mu(\mathbf{r})/\varepsilon_F(\mathbf{r})=0.358$ of our 
numerical calculations \cite{Haussmann3} we obtain the result $r_{TF}/R_{TF}=0.773$. 
This value agrees very well with the most recent field theoretic result $\xi=0.367(9)$ 
for the Bertsch parameter, obtained from a Borel resummation of an expansion around 
the upper critical dimension four, carried out to three loop order 
\cite{Arnold07,Nussinov06}. It is somewhat smaller, however, than the result 
$r_{TF}/R_{TF}=0.80$ found by using the Bertsch parameter $\xi=0.42(1)$ that follows 
from variational Monte Carlo calculations \cite{Carlson03,Astra04} or from a theory 
that includes the Gaussian fluctuations around the BCS mean field, extended to arbitrary 
coupling, where $\xi=0.40$ \cite{HLD06,DSR08}. 

Next we insert the local internal energy per particle $u(\mathbf{r})$, the potential 
$V(\mathbf{r})=\frac{1}{2} m\omega^2r^2$, and the local particle density 
$n(\mathbf{r})$ into Eqs.\ \eqref{Eq:potential_energy}-\eqref{Eq:total_energy} in 
order to calculate the energies of the unitary Fermi gas in the harmonic trap. 
We thus obtain the dimensionless ratios
\begin{eqnarray}
2\, \frac{E_\mathrm{pot}}{N E_F} &=& 
\frac{3}{4}\, \frac{E_F}{\varepsilon_F(\mathbf{0})} =
\frac{3}{4} \left[ \frac{\mu(\mathbf{0})}{\varepsilon_F(\mathbf{0})} \right]^{1/2} 
\nonumber\\ 
&=& 0.449 \ ,
\label{Eq:potential_energy_3} \\
2\, \frac{U}{N E_F} &=& \frac{3}{4}\, \frac{u(\mathbf{0})}{E_F} =
\frac{3}{4}\, \frac{u(\mathbf{0})}{\varepsilon_F(\mathbf{0})}
\left[ \frac{\mu(\mathbf{0})}{\varepsilon_F(\mathbf{0})} \right]^{-1/2} \nonumber\\
&=& 0.440 \ ,
\label{Eq:internal_energy_3} \\
\frac{E}{N E_F} &=& \frac{U+E_\mathrm{pot}}{N E_F} = 0.444 \ .
\label{Eq:total_energy_3}
\end{eqnarray}
The explicit numbers are obtained by inserting the ratios
$\mu(\mathbf{r})/\varepsilon_F(\mathbf{r})=0.358$ and
$u(\mathbf{r})/\varepsilon_F(\mathbf{r})=0.210$ of our numerical calculation 
\cite{Haussmann3}.

For a harmonic potential $V(\mathbf{r})$ it is well known that the internal energy
$U$ and the potential energy $E_\mathrm{pot}$ are related to each other by the
virial theorem $U=E_\mathrm{pot}$. As shown by Thomas \textit{et al.}\ 
\cite{Thomas05}, this theorem also holds for the interacting Fermi gas
in the unitary regime. As a consequence, the results of Eqs.\ 
\eqref{Eq:potential_energy_3}-\eqref{Eq:total_energy_3} should be equal. By using 
the dimensionless ratios of the homogeneous system \eqref{Eq:dimensionless_ratios} 
we can express the results in terms of the Bertsch parameter $\xi$ according to
\begin{equation}
\frac{E}{N E_F} = 2\, \frac{U}{N E_F} = 2\, \frac{E_\mathrm{pot}}{N E_F} =
\frac{3}{4}\,\xi^{1/2} \ .
\label{Eq:energies_4}
\end{equation}
In practice, our resulting numbers differ by about $2.0\%$. This 
difference is related to the fact, that the exact relation 
$u(\mathbf{r})/\mu(\mathbf{r})=3/5$ for the unitary gas is satisfied only
with an error of $2.0\%$ in our theory. 

The result \eqref{Eq:energies_4} for the ground state energy in the trap
is based on using LDA and provides the exact leading order contribution in
the limit $N\to\infty$. The question of how large the subleading corrections to
this result are has been addressed by Son and Wingate \cite{Son06}. Using 
a gradient expansion of the effective field theory describing the low energy
physics of the fermionic superfluid at unitarity, they have determined 
the $q^2$-corrections to the density response, which is equal to the 
uniform compressibility $\partial n/\partial\mu$ at $q=0$. These corrections
give rise to an additional contribution \cite{Son06}
\begin{equation}
\begin{split}
\frac{E}{N E_F} = \frac{3}{4} \,\xi^{1/2} \Bigl[ 1 \ + \ &4\pi^2 (2\xi)^{1/2} 
\Bigl( \frac{9}{2} c_2 - c_1 \Bigr) \\
&\times \frac{\omega_x^2+\omega_y^2+\omega_z^2}{\omega^2} 
(3N)^{-2/3} + \ldots \Bigr]  \ .
\end{split}
\label{Eq:energies_5}
\end{equation}
to the ground state energy, which contains the two dimensionless coefficients 
$c_1$ and $c_2$ which appear beyond the leading coefficient $\xi$ of 
the uniform system in an expansion up to second order in gradients. The leading 
correction to LDA is thus of relative order $(3N)^{-2/3}$  (formally it is 
$\sim\hbar^2$) and describes a {\it curvature} instead of the naively expected 
{\it surface} correction, which would scale like $N^{-1/3}$. The absence of a 
surface correction also apppears for an ideal Fermi gas and is a peculiarity of 
the harmonic confinement \cite{Brack_1}. The prefactor of the $(3N)^{-2/3}$ 
correction in \eqref{Eq:energies_5} has been determined from an expansion around 
dimension four by Rupak and Sch\"afer \cite{Rupak08} and is $2.41$ for an isotropic 
trap. For experimentally relevant particle numbers $N \approx 1.3 \times 10^5$, the 
beyond LDA relative corrections to the ground-state energy are therefore only 
around $4.5 \times 10^{-4}$ and thus are clearly below the present experimental 
accuracy.

The final results of this section are the Thomas-Fermi radius $r_{TF}/R_{TF}=0.773$ 
and the total energy $E/N E_F=0.444$ of the unitary Fermi gas at zero temperature in 
a harmonic trap, which depend only on the Bertsch parameter $\xi=0.358$ of the 
homogeneous system by Eqs.\ \eqref{Eq:Thomas_Fermi_radius} and \eqref{Eq:energies_4}.
These results will be compared with experiments and other theories in the next 
section, where we discuss the situation at finite temperatures.

\section{Unitary regime for nonzero temperatures}
\label{section_4}
For nonzero temperatures, the local thermodynamic quantities are space dependent 
via the scale factors $k_F(\mathbf{r})$, $\varepsilon_F(\mathbf{r})$, and also
via the dimensionless temperature $\theta(\mathbf{r})=k_B T/\varepsilon_F(\mathbf{r})$. 
In the unitary regime the dimensionless interaction 
$v(\mathbf{r})=1/k_F(\mathbf{r}) a=0$ is constant. Since the effective 
temperature increases towards the edge of the cloud, the local 
particle density $n(\mathbf{r})$ does not follow the simple 
$[ 1 - r^2 / r_{TF}^2 ]^{3/2}$ law valid at $T=0$ and Eq.\ \eqref{Eq:chempot_condition} 
must be solved numerically in order to obtain the detailed density profile 
$n(\mathbf{r})=n(r)$ as a function of the weighted radial coordinate $r$. 

\begin{figure}[b]
\includegraphics[width=0.44\textwidth,clip]{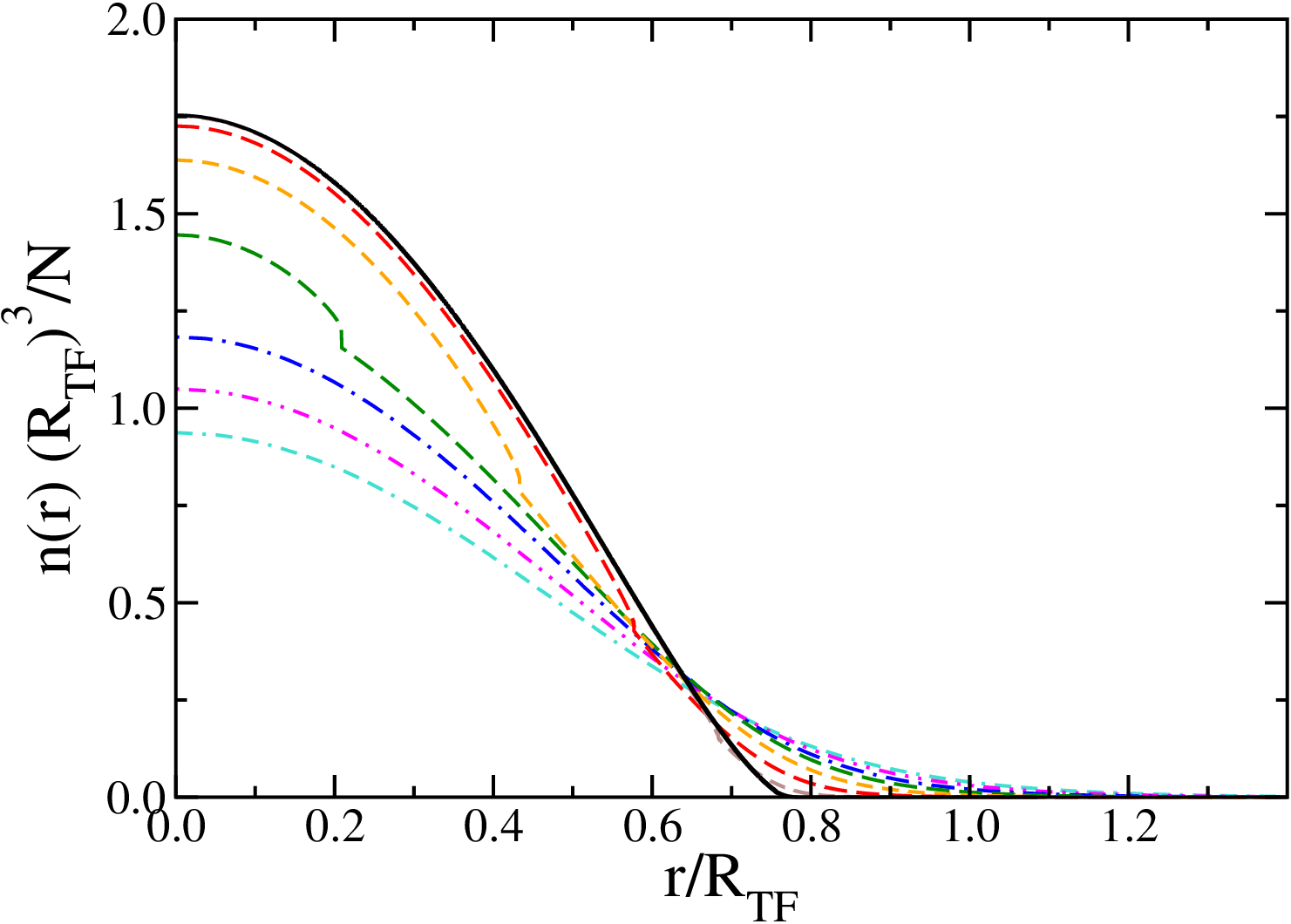}
\caption{(Color) The local particle density $n(r)$ of the unitary Fermi gas 
in a harmonic trap as a function of the radius $r$ for several temperatures 
$\theta(\mathbf{0})= k_B T/\varepsilon_F(\mathbf{0})=0.000$ (black solid), 
$0.032$ (brown short dashed), $0.065$ (red long dashed), $0.100$ (orange short dashed), 
$0.141$ (green long dashed), $0.192$ (blue dot-dashed), 
$0.252$ (magenta doubledot-dashed), and $0.313$ (turquoise dot-doubledashed). 
The length scale is $R_{TF}=(24N)^{1/6} (\hbar/m\omega)^{1/2}$.}
\label{Fig:density_profiles}
\end{figure}

Using the results of our previous numerical calculations \cite{Haussmann3} for the 
homogeneous system, we obtain the density profiles $n(r)$ for several values of the 
temperature $T$ which are shown in Fig.\ \ref{Fig:density_profiles}. The black solid 
line is the density profile for zero temperature. It is nearly identical 
to the expected zero temperature Thomas-Fermi profile. Differences are due to the
limited numerical accuracy, which are, however, much smaller than the thickness of 
the line. At zero temperature, the Fermi gas is superfluid in the whole trap. The 
colored and dot-dashed lines represent the density profile for nonzero temperatures. 
For very low temperatures, changes occur only close to the surface of the cloud 
where $r/R_{TF}\approx r_{TF}/R_{TF}=0.77$. For this reason, the brown short dashed 
line is visible only for $0.65 \lesssim r/R_{TF} \lesssim 0.85$ while otherwise it 
is nearly on top of the black solid line. The red long dashed line is still very 
close to the black solid line. At a position in space $\mathbf{r}$ the Fermi 
gas may be normal fluid or superfluid if the local dimensionless temperature
$\theta(\mathbf{r})= k_B T/\varepsilon_F(\mathbf{r})$ is above or below the 
critical value $\theta_c\approx 0.16$. For the brown, red, orange, and green dashed 
lines there exists a respective weighted radius $r_c$ so that $\theta(r_c)=\theta_c$. 
In these cases the Fermi gas is superfluid in the inner region of the trap where 
$r<r_c$ and normal fluid in the outer region where $r>r_c$. For the blue,
magenta, and turquoise dot-dashed lines the dimensionless temperature is 
$\theta(\mathbf{r})>\theta_c$ for all positions in space $\mathbf{r}$ so that
the Fermi gas is normal fluid in the whole trap.

Since $\theta(\mathbf{r})$ has its lowest value for $\mathbf{r}=\mathbf{0}$, the 
dimensionless temperature at the center of the trap $\theta(\mathbf{0})$ determines
the superfluid transition of the confined Fermi gas. For $\theta(\mathbf{0})>\theta_c$ 
the Fermi gas is completely normal fluid, while for $\theta(\mathbf{0})<\theta_c$ 
there exists a superfluid region close to the center.

In a homogeneous gas, the normal to superfluid transition is a continuous 
phase transition of the 3D XY type along the complete BCS to BEC crossover, 
because the broken symmetry is that associated with a complex scalar
order parameter. By contrast, our approach \cite{Haussmann3} predicts a weak 
first-order superfluid transition because the superfluid phase of the Luttinger-Ward 
theory does not smoothly connect with the normal-fluid phase at a single critical 
temperature $\theta_c = k_B T_c / \varepsilon_F$. As a result, there are two 
slightly different critical temperatures 
$\theta_{c,\mathrm{upper}}$ and $\theta_{c,\mathrm{lower}}$. The upper value 
$\theta_{c,\mathrm{upper}}$ is defined by the condition that for 
temperatures above this value the superfluid order parameter vanishes, 
while the lower value $\theta_{c,\mathrm{lower}}$ is defined by the 
temperature below which the normal-fluid phase is no longer stable.  
Fortunately, the difference between both temperatures, which should 
vanish in an exact theory, is rather small over essentially the whole 
BCS to BEC crossover. In particular, at unitarity, the upper and lower 
values for $\theta_c$ are $0.1604$ and $01506$, which is within the 
present numerical uncertainties in the determination of the 
critical temperature of the unitary gas.  
Indeed, our critical temperature agrees very well with 
the most precise calculations of $\theta_c$ so far by 
Quantum Monte Carlo calculations, which give $\theta_c=0.152(7)$
for the uniform gas at unitarity \cite{Burovski06}, a value that 
has been confirmed very recently \cite{Burovski08}.

The existence of two different critical temperatures
leads to a multivaluedness of thermodynamic quantities, which 
is an artifact of the first order nature of the superfluid transition 
within our theory. In order to avoid multivalued local density 
or entropy profiles, we have connected the normal and
superfluid branches with a kink at the point, where they are 
closest, thus providing an optimal approximation to the exact
continuous profiles in a theory which properly accounts for the continuous 
nature of the transition in the infinite system. 
This point is related to the upper value of the dimensionless 
critical temperature $\theta_{c,\mathrm{upper}}$. Hence, 
in Fig.\ \ref{Fig:density_profiles} the brown, red, orange, 
and green dashed lines have kinks located at 
$r_{c,\mathrm{upper}} / R_{TF} = 0.684$, $0.577$, $0.433$, $0.209$, 
respectively. The inner, superfluid branches of these lines
show a bulge in the center of the trap, which is appreciable, in particular 
for the green curve. The formation of a small bulge upon condensation,
which first appears near the trap center, has indeed been observed 
experimentally \cite{KZ08}, although the effect in axially integrated profiles
is rather small. 

Previous theoretical results for the  density profile have been obtained by 
Bulgac \textit{et al.}\ \cite{Bulgac07} within a Monte-Carlo approach. For 
zero temperature, their result is very close to our black solid line in 
Fig.\ \ref{Fig:density_profiles}. They obtain the Thomas-Fermi radius 
$r_{TF}/R_{TF}=0.81$ \cite{Drut_2}. This is larger than our value $0.77$ 
because of the larger value $\xi=0.43$ of the Bertsch parameter. For nonzero 
temperatures (the same as those in Fig.\ \ref{Fig:density_profiles}), 
Bulgac \textit{et al.}\ obtain single valued density profiles which agree 
qualitatively with our results but differ somewhat quantitatively. They also 
observe a superfluid bulge which, however, is smaller.

\begin{figure}[t]
\includegraphics[width=0.44\textwidth,clip]{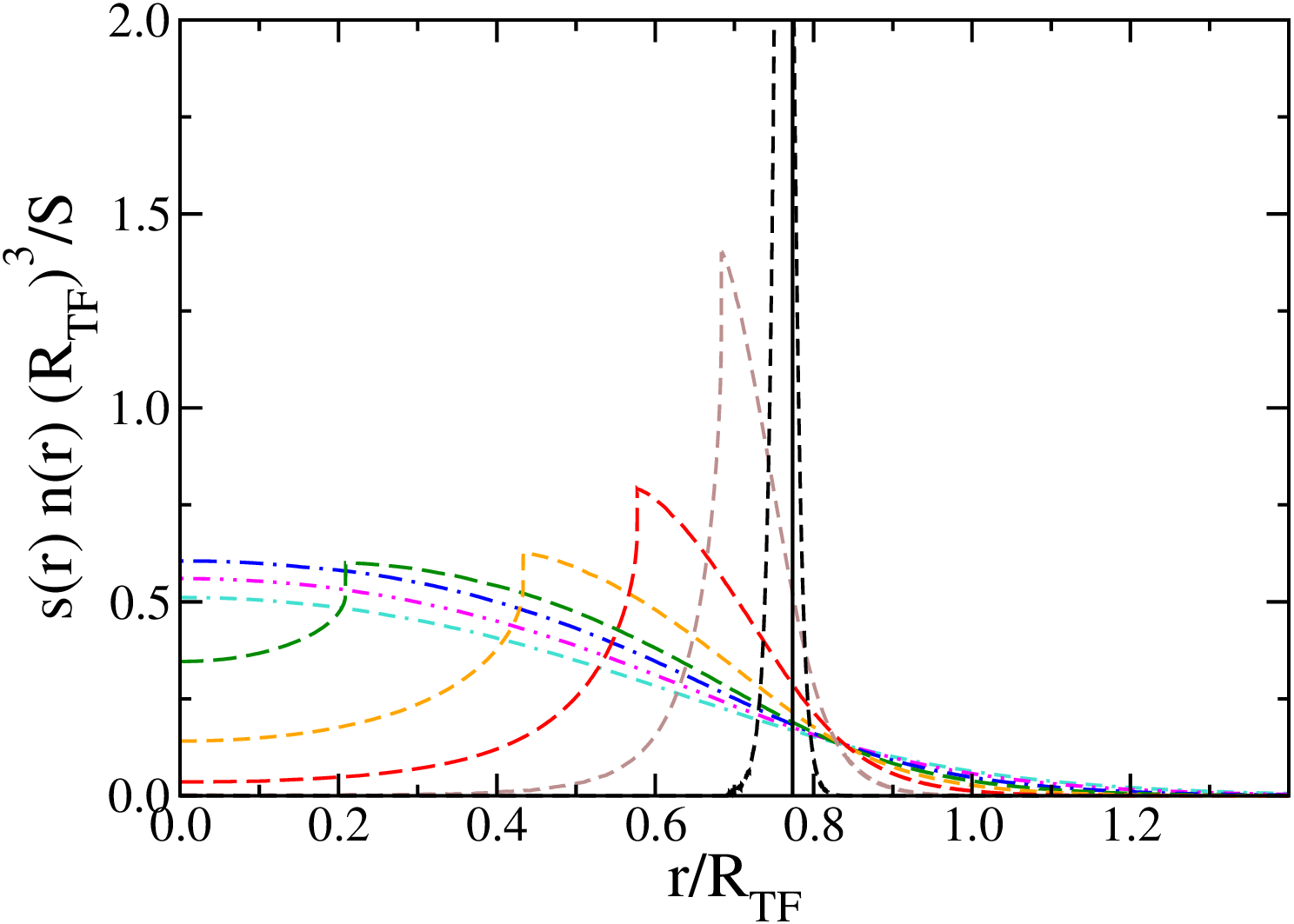}
\caption{(Color) The local entropy density $s(r) n(r)$ of the unitary 
Fermi gas in a harmonic trap as a function of the radius $r$ for several 
temperatures. The curves are related to those in Fig.\ \ref{Fig:density_profiles}. 
An additional curve is included for 
$\theta(\mathbf{0})= k_B T/\varepsilon_F(\mathbf{0})=0.0065$ (black short dashed).}
\label{Fig:entropy_profiles}
\end{figure}

A very interesting quantity is the local entropy per particle $s(r)$. From our 
numerical calculations \cite{Haussmann3} of the homogeneous system, and within LDA, 
the resulting profiles $s(r) n(r)$ of the local entropy per \textit{volume} are 
shown in Fig.\ \ref{Fig:entropy_profiles} for several values of the temperature $T$. 
The colors of the curves are related to those in Fig.\ \ref{Fig:density_profiles}.
For high temperatures where the Fermi gas is completely normal fluid (blue, magenta,
and turquoise dot-dashed curves), the entropy density is distributed over the whole 
trap with a maximum at the center. For low temperatures (orange, red, and brown 
dashed curves) the center part is superfluid but the outer part is normal fluid. 
Consequently, in these cases the entropy density is minimum in the center of the 
trap but maximum at a nonzero radius. For very low temperatures (black dashed line), 
the main contribution of the entropy is located close to the surface of the atom 
cloud. The width of this surface layer decreases for decreasing temperature and 
eventually shrinks to zero in the zero temperature limit (black solid line).

In Fig.\ \ref{Fig:entropy_profiles} we have again eliminated multivalued 
regions which are an artifact of the first-order nature of the transition in our 
theory for the homogeneous system by using the same criterion as in the density
profiles of  Fig.\ \ref{Fig:density_profiles}. Indeed, since the normal to superfluid
transition is not associated with a latent heat, the local entropy density will be 
single valued. An interesting observation that  is evident from
Fig.\ \ref{Fig:entropy_profiles}, is that sufficiently below the onset of 
superfluidity in a trapped Fermi gas, most of the entropy is located in the 
normal region near the cloud edge. This observation suggests an 
efficient way to lower the entropy further by removing atoms in the boundary 
layer and simultaneously readjusting the trapping potential such that the now 
smaller system has a radius close to that beyond which atoms have 
been removed. This idea is - of course - in the same spirit than the standard 
evaporation cooling in the normal state \cite{Ketterle97}, however it is much 
more efficient. Indeed, consider starting with a dimensionless temperature  
$\theta(\mathbf{0})= 0.065$ (red long dashed curve) or - equivalently - an entropy 
per particle $S/N k_B= 0.433$ which is close to that reached in current experiments. 
Removing about $42$ percent of the particles in the shell beyond $r=0.5\,R_{TF}$, 
will lower the entropy per particle by a factor $8.8$ to $S/N k_B= 0.049$ and the 
dimensionless temperature by a factor $2.6$ to $\theta(\mathbf{0})= 0.025$.
In turn, for an initial temperature $\theta(\mathbf{0})= 0.192$ (blue dot-dashed curve)
where the cloud is a normal gas, removing the same amount of particles 
will reduce the entropy only by a factor $1.6$. Removing atoms  in 
the outer shell repeatedly thus provides an effective tool to reach very low 
entropies and temperatures. In practice, for atoms confined in an optical dipole trap,
this may be achieved by lowering the depth of the optical trap, as was done e.g.\ 
previously in experiments realizing a condensate of fermionic dimers \cite{Jochim03}.

\begin{figure}
\includegraphics[width=0.44\textwidth,clip]{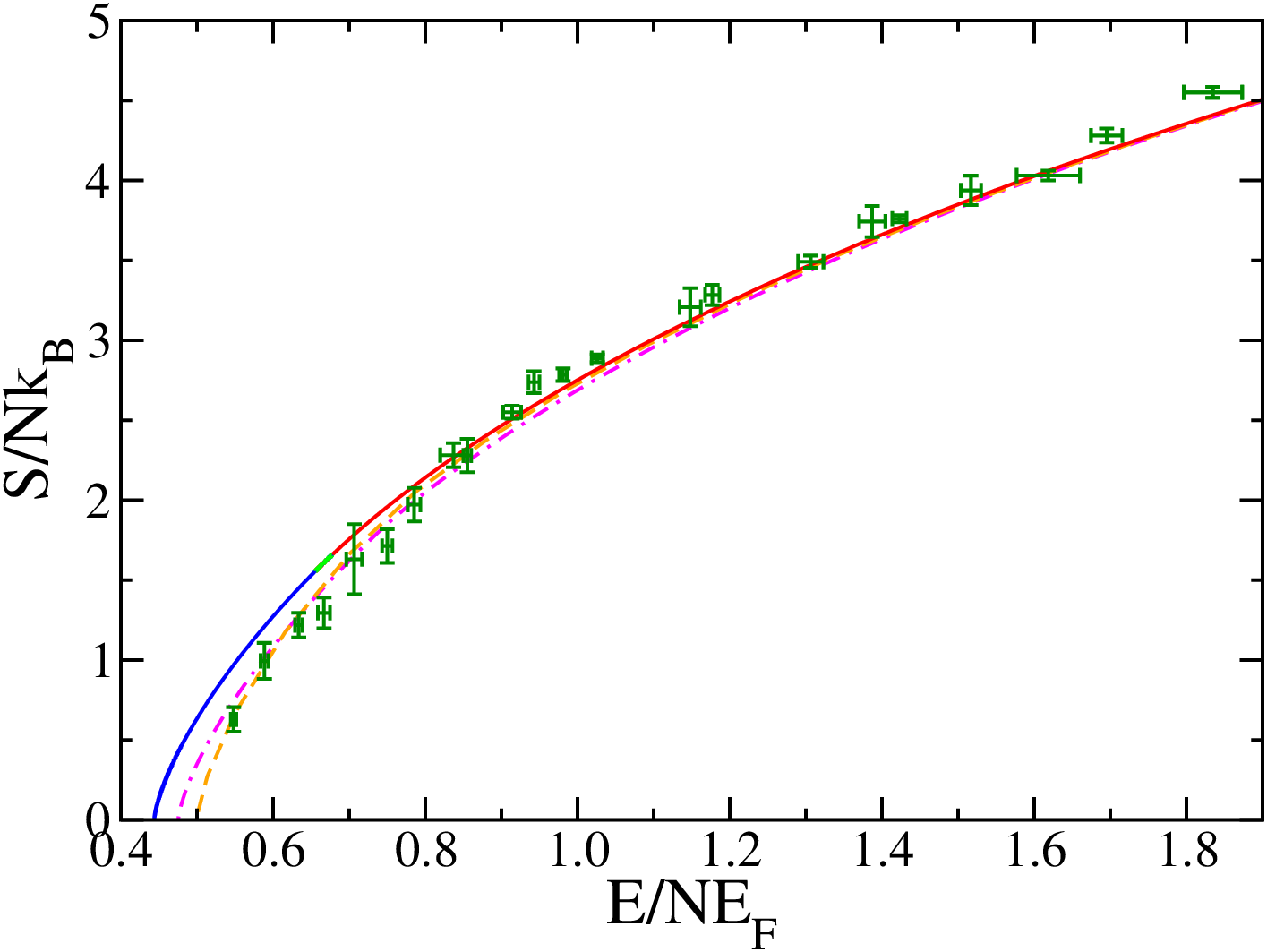}
\caption{(Color) The entropy $S$ as a function of the total energy $E$ for 
the unitary Fermi gas in a harmonic trap: present theory (blue-green-red solid), 
NSR theory of Hu \textit{et al.}\ \cite{HLD06,HDL07,HLD08} (magenta dot-dashed)
Monte-Carlo simulation of Bulgac \textit{et al.}\ \cite{Bulgac07} (orange dashed), 
and experimental data of Luo \textit{et al.}\ \cite{Luo07} (green data points).}
\label{Fig:entropy}
\end{figure}

From the local profiles $n(\mathbf{r})$, $u(\mathbf{r})$, and $s(\mathbf{r})$, it 
is straightforward to calculate the the total Energy $E$ and the global entropy $S$ 
by evaluating the integrals in Eqs.\ \eqref{Eq:total_energy} and \eqref{Eq:entropy}. 
In particular, we may eliminate the local value $\theta(0)$ of the dimensionless
temperature in the trap center, to obtain the function $S=S(E)$. 
Our result for the ultracold Fermi gas in the unitary regime is shown in 
Fig.\ \ref{Fig:entropy} as blue-green-red solid line. To distinguish the
superfluid and the normal-fluid parts of the curve, the solid line is 
shown in blue or red color, respectively. Apparently, this curve is 
continuous and there is no particular feature, which indicates the superfluid 
transition. In fact, since $dS/dE$ is just the inverse temperature, this behavior 
is expected not only in a trap, where none of the thermodynamic functions exhibits 
a singularity, but even in the homogeneous gas, because the superfluid transition is
continuous. As mentioned before, however, our theory predicts a weak first-order 
transition. The solid line is therefore multivalued in the intervals 
$0.656<E/N E_F<0.677$ and $1.56<S/N k_B<1.66$, which is indicated by the green 
section of the line. Within these intervals the superfluid transition is located. 
In practice, evidently, the multivaluedness is so tiny that it can hardly be seen in 
Fig.\ \ref{Fig:entropy}. In the multivalued region the blue, green, and red branch 
of the solid line therefore lie on top of teach other.

Bulgac \textit{et al.}\ have obtained $S(E)$ from their Monte-Carlo calculation 
\cite{Bulgac07}, which, in Fig.\ \ref{Fig:entropy}, is shown as orange dashed line. 
Clearly, the agreement with our theory is nearly perfect for high temperatures well 
above the superfluid transition. 
However, for lower temperatures close to and below the transition, the results
differ. The deviations are largest for zero temperature where $S=0$. They are 
essentially due to the fact that the ground-state energy $E_0$ in both approaches 
differ. Indeed, from Eq.\ \eqref{Eq:total_energy_3} we obtain $E_0/N E_F = 0.444$
while the corresponding value of Bulgac \textit{et al.}\ $E_0/N E_F = 0.50$ is 
larger because of the larger value $\xi=0.43$ of the Bertsch parameter. 
Apart from the deviations in the limit of zero temperature, 
the theories also differ in their predictions
of the behavior near the superfluid transition temperature. In particular, 
Bulgac \textit{et al.}\ \cite{Bulgac07} have calculated the critical 
values $E_c/N E_F = 0.50+0.32=0.82$ and $S_c/N k_B= 2.15$ at the superfluid 
transition. These results are considerably larger than our predictions 
$0.656<E_c/N E_F<0.677$ and $1.56<S_c/N k_B<1.66$, whose uncertainty
is due to the multivaluedness near $T_c$. As  a result, the value of 
the critical temperature $k_B T_c/E_F=0.27$ of the unitary Fermi gas in a trap
in the theory of Ref.\ \onlinecite{Bulgac07}  is considerably larger than our 
prediction $0.207< k_B T_c/E_F< 0.220$. 

Now, as pointed out above, our result for the critical temperature 
of the uniform gas agrees rather well with the most 
precise numerical calculations of this quantity \cite{Burovski06,Burovski08}. 
In addition, it is also consistent with the recent calculations of Bulgac 
\textit{et al.}\ \cite{Bulgac08}, which indicate that 
$\theta_c = k_B T_c/\varepsilon_F $ is less or equal to $0.15(1)$, again in very 
good agreement with our numbers. This underlines that our self-consistent, 
conserving theory of the BCS-BEC crossover \cite{Haussmann3} provides a 
quantitatively reliable description of the thermodynamics of a balanced Fermi gas 
near unitarity, despite the problems with the first order nature of the transition 
in our approach and the absence of a small expansion parameter. Within LDA, which 
provides a rather accurate description in the relevant regime of particle numbers
$N\approx 10^5$, our critical temperature $k_B T_c= 0.21(1)\, E_F$ for the trapped
gas is therefore expected to be close to the exact  result. This is consistent with 
a very recent analysis of the experimental data of Ref.\ \onlinecite{Luo07}, which 
indicates a critical temperature very close to this value \cite{Thomas08}.
Previous, much higher values of the critical temperature for both the 
homogeneous or the trapped gas that were obtained in different extensions
of the theory by Nozi\`eres and Schmitt-Rink \cite{Perali04,HLD06}
and also in numerical calculations \cite{Bulgac06,Ceperley07} are clearly  
ruled out.

An alternative approach to the BCS to BEC crossover problem has been developed by 
Hu \textit{et al.}\ \cite{HLD06} for the homogeneous system and applied to the 
ultracold Fermi gas in a trap \cite{HDL07,HLD08}. This theory is an extension of the 
approach by Nozi\`eres and Schmitt-Rink (NSR) \cite{NSR} to the superfluid region at 
low temperatures. While the order parameter $\Delta$ is determined by the standard 
mean-field gap equation, the chemical potential $\mu$ is calculated by a 
particle-density equation, which includes condensed and noncondensed bound pairs.
The extended NSR approach is in fact  a limiting case of
our theory in which the full, self-consistently determined
Green functions are replaced by their zeroth order form obtained in BCS theory.
Hu \textit{et al.}\ \cite{HDL07,HLD08} have calculated $S=S(E)$ and obtain a result 
which is shown as magenta dot-dashed line in Fig.\ \ref{Fig:entropy}. Again, the 
agreement is very good for high temperatures in the normal fluid region while, 
however, for low temperatures in the superfluid region there are deviations. Hu 
\textit{et al.}\ \cite{HDL07,HLD08} obtain the ground-state energy $E_0/N E_F = 0.47$ 
which is related to the Bertsch parameter $\xi=0.40$. Moreover, similar to the theory
of Bulgac \textit{et al.}\ \cite{Bulgac07} their critical temperature 
$k_B T_c= 0.25\, E_F$ and entropy $S_c\simeq 2.2\, Nk_B$ are considerably larger 
than our values.

More recently, Hu \textit{et al.}\ \cite{HLD08} have published a self-consistent 
result for the entropy $S(E)$ which they call ``$GG$ approximation''. This method 
is equivalent to our approach for the homogeneous system \cite{Haussmann3} and gives 
results that are nearly identical to the prediction of our present theory including 
the critical temperature $k_B T_c = 0.21\, E_F$.

While in the experimental setup the correct trap potential is Gaussian, all the curves 
shown in Fig.\ \ref{Fig:entropy} are calculated for the harmonic potential 
\eqref{Eq:harmonic_potential}. Slight differences would occur in the high-temperature 
regime where the energies $E$ and the entropies $S$ are large. However, it is important 
that all curves are calculated for the same potential so that they all converge to a 
single line in the high-temperature limit.

The entropy versus the total energy has been measured experimentally for ultracold 
$^6\mathrm{Li}$ atoms in an optical trap by Luo \textit{et al.}\ \cite{Luo07}. 
The data are shown as green points with horizontal and vertical error bars. Clearly 
the data agree with all theories in the high-temperature regime. The slightly 
larger entropy values of the experimental data for large energies may be due to the 
fact that the experiment is performed for a Gaussian potential while the theoretical 
curves are calculated for a harmonic potential. Apparently, for low 
temperatures and low entropies the experimental data agree better with the theories 
of Bulgac \textit{et al.}\ \cite{Luo07} and of Hu \textit{et al.}\ \cite{HDL07,HLD08} 
than with our theory. In particular, an extrapolation to zero entropy gives a 
ground-state energy $E_0/N E_F = 0.53$, considerably higher than our value $0.444$.
It is difficult, however, to quantify the error involved in such an extrapolation 
because the determination of the entropy from a comparison with its ideal Fermi gas 
limit reached after an adiabatic ramp to magnetic fields $B=1200\,\mathrm{G}$ far on 
the BCS side of the crossover becomes increasingly difficult as $S$ approaches zero. 
As pointed out above, a quite sensitive parameter which distinguishes previous 
theories from our present one is the critical temperature of the superfluid transition
and the associated value of the entropy and energy. Unfortunately, it is difficult to 
determine the critical temperature and thus also the corresponding value of the 
entropy from measurements of $S(E)$. Based on the excellent agreement of our 
value for $T_c$ and the Bertsch parameter $\xi$ with the most precise numerical or 
field-theoretical results and the fact that all thermodynamic relations are obeyed 
at the few percent level, it is likely that our present theory gives a reliable 
description of the thermodynamics of the trapped unitary gas, despite the fact that, 
superficially, the agreement with the data shown in Fig.\ \ref{Fig:entropy} is not as 
good as those of previous theories.

\section{Thermometry}
\label{section_5}

The most important parameter for thermodynamic properties is - of course - 
the temperature, which unfortunately cannot be measured directly. 
In principle, this is possible from the density profile $n(\mathbf{r})$ which 
changes as a function of temperature. For balanced gases, however, this method 
is not very reliable. Indeed, for low temperatures, the density converges to the 
simple Thomas-Fermi profile. As shown in Fig.\ \ref{Fig:density_profiles}, the 
deviations from such a profile at finite temperatures are extremely small for the 
inner part of the atom cloud (see the red and brown low-temperature curves, 
which are nearly identical to the black zero-temperature curve). 
Only close to the surface $r/R_{TF} \approx 0.77$, small variations with the 
temperature are observed. Therefore, the signal to noise ratio will be small 
in the interesting regime below $T/T_F\approx 0.1$.

A quantity which is much more sensitive to temperature is the 
entropy density, shown in Fig.\ \ref{Fig:entropy_profiles}. At low 
temperatures, it  is peaked near the surface of the atom cloud. Unfortunately, 
the local entropy density $s(\mathbf{r}) n(\mathbf{r})$ is not accessible 
experimentally. We therefore consider the total entropy $S$, which has been 
measured by \cite{Luo07}, as discussed above.

\begin{figure}
\includegraphics[width=0.44\textwidth,clip]{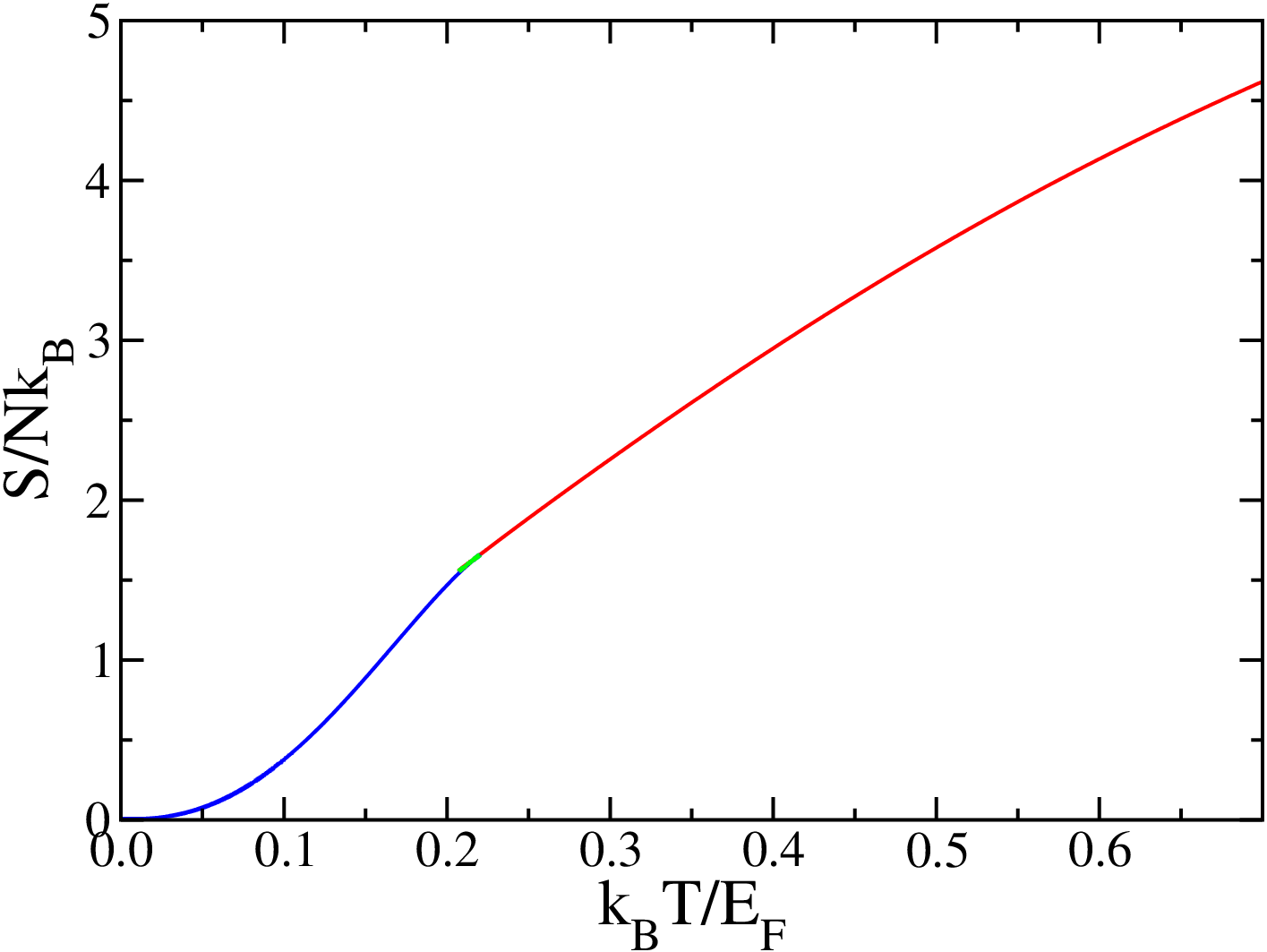}
\caption{(Color) The entropy $S$ as a function of the temperature $T$ for 
the unitary Fermi gas in a harmonic trap. The blue, red, and green sections 
represent the superfluid, normal-fluid and multivalued branches of the curve, 
respectively.}
\label{Fig:entropy-temperature}
\end{figure}

In Fig.\ \ref{Fig:entropy-temperature}, we plot the total entropy $S$ in a trap 
as a function for the temperature $T$ in units of the Fermi temperature 
$T_F= E_F/k_B= (3N)^{1/3}\hbar\omega/k_B$ of the trapped non-interacting gas. 
As in Fig.\ \ref{Fig:entropy}, we indicate the superfluid and the normal-fluid 
region by blue and red color of the curve, respectively. Again, a tiny multivalued 
region is observed close to the superfluid transition which is indicated by 
the green section of the line. However, the three branches in the multivalued 
region are nearly on top of each other. Consequently, the entropy $S=S(T)$ is  
effectively continuous at the superfluid transition as expected. Using the curve 
in Fig.\ \ref{Fig:entropy-temperature}, the measurements of the entropy by Luo 
\textit{et al.}\ \cite{Luo07} therefore allow to reliably infer the related 
temperatures $T$.

Alternatively, we may consider the excess internal energy density $[u(r)-u_0(r)] n(r)$ 
where $u_0(r)$ is the ground-state internal energy per particle at zero temperature. 
The related profiles are similar to those shown in Fig.\ \ref{Fig:entropy_profiles}. 
For low temperatures the excess internal energy density is peaked close to the surface 
of the atom cloud. Integrating over the space we obtain the excess internal energy 
$U-U_0$ which is related to the excess total energy by the virial theorem 
$E-E_0= 2 (U-U_0)$. Moreover, the total energy $E$ is related to the mean 
square radius of the atom cloud $\langle \mathbf{r}^2 \rangle$ according to 
$E=2 E_\mathrm{pot}=N m\omega^2 \langle \mathbf{r}^2 \rangle$. Since 
$\langle \mathbf{r}^2 \rangle$ can be measured in a model-independent way 
\cite{Thomas05,Luo07}, the total energy $E=E(T)$ as a function of the 
temperature $T$ provides an alternative method to determine the temperature 
of the interacting Fermi gas. We obtain a curve which is similar like 
Fig.\ \ref{Fig:entropy-temperature}. However, the major drawback of this method is 
that it requires knowledge of the ground state energy $E_0$, which must be determined 
with sufficient accuracy. As evident from Fig.\ \ref{Fig:entropy}, there are 
significant discrepancies in the ground state energy $E_0$ for different experiments 
and theories.

\section{Low-energy collective modes}
\label{section_6}

In the superfluid regime at very low temperatures the fermionic quasiparticles are 
frozen out and do not contribute to thermodynamic quantities because they have an 
energy gap which is related to the binding energy of the Cooper pairs. However, the 
spontaneous symmetry breaking implies a gapless Goldstone mode which is the 
Bogoliubov-Anderson mode. This mode propagates with a constant velocity $c$ like 
phonons. For this reason, the low temperature behavior of the entropy density and the 
internal energy density is ruled by the well known Stefan-Boltzmann formulas
\begin{eqnarray}
s(\mathbf{r})\, n(\mathbf{r}) &=& 
\frac{8}{3}\, \frac{\sigma(\mathbf{r})}{c(\mathbf{r})}\, T^3 \ , 
\label{Eq:stefan_boltzmann_entropy} \\
\left[ u(\mathbf{r}) - u_0(\mathbf{r}) \right] n(\mathbf{r}) &=& 
2\, \frac{\sigma(\mathbf{r})}{c(\mathbf{r})}\, T^4 
\label{Eq:stefan_boltzmann_energy}
\end{eqnarray}
for phonons with one polarization degree of freedom where 
$\sigma(\mathbf{r})=(\pi^2 k_B^4)/(60\hbar^3 [c(\mathbf{r})]^2)$ is the Stefan-Boltz\-mann 
factor. Since the sound velocity $c(\mathbf{r})$ depends on the local particle density, 
$\sigma(\mathbf{r})$ and $c(\mathbf{r})$ are space-dependent parameters. Eliminating the 
temperature we obtain a relation between the local entropy per particle $s(\mathbf{r})$ 
and the local internal energy per particle $u(\mathbf{r})$, which can be written in 
the form 
\begin{eqnarray}
\frac{s(\mathbf{r})}{k_B} = \frac{2\,\pi}{3\times 5^{1/4}} 
\left[ \frac{v_F(\mathbf{r})}{c(\mathbf{r})}
\ \frac{u(\mathbf{r})-u_0(\mathbf{r})}{\varepsilon_F(\mathbf{r})} \right]^{3/4} \ .
\label{Eq:local_asymptotic_formula}
\end{eqnarray}
Here $v_F(\mathbf{r})= \hbar k_F(\mathbf{r})/m$ is the local Fermi velocity, 
and $u_0(\mathbf{r})$ is the ground-state energy per particle. The ratio 
$c(\mathbf{r}) / v_F(\mathbf{r})= c/v_F= (\xi/3)^{1/2}= 0.345$ is constant and 
related to the Bertsch parameter $\xi=0.358$.

In our previous publication \cite{Haussmann3} we have calculated all thermodynamic
quantities for the homogeneous system. Hence, also the function $s=s(u)$ is 
available, so that we can check the asymptotic formula 
\eqref{Eq:local_asymptotic_formula} for the homogeneous system. It turns out that
the resulting exponent at low energy is indeed equal to $3/4$. However the amplitude,
which is determined fully by the sound velocity $c$, gives $c/v_F=0.7$ for the unitary 
Fermi gas. This is about a factor of two larger than the expected value 
$c/v_F=(\xi/3)^{1/2}$ from the Bertsch parameter. This discrepancy indicates, that 
the accuracy of our theory at very low temperatures is not sufficient to extract a 
reliable value of the sound velocity from the entropy.

In the trapped case, unfortunately, asymptotic formulas like 
\eqref{Eq:stefan_boltzmann_entropy}-\eqref{Eq:local_asymptotic_formula} 
do not hold for the total entropy $S$ and the total energy $E$. To see this, 
we integrate Eqs.\ \eqref{Eq:stefan_boltzmann_entropy} 
and \eqref{Eq:stefan_boltzmann_energy} over the whole space. Using Eqs.\ 
\eqref{Eq:entropy} and \eqref{Eq:internal_energy} we obtain well defined results 
$S$ and $U-U_0=\frac{1}{2}(E-E_0)$ for the left hand sides, respectively. 
However, while the temperature $T$ is constant, the space dependence on the 
right-hand sides arise from the factor 
$\sigma(\mathbf{r}) / c(\mathbf{r}) \sim [c(\mathbf{r})]^{-3} 
\sim [v_F(\mathbf{r})]^{-3} \sim [n(\mathbf{r})]^{-1}$. This factor is minimum at 
the center of the trap but maximum at the surface of the atom cloud. Thus, the 
main contribution of the integral arises from the surface of the atom cloud (see 
also Fig.\ \ref{Fig:entropy_profiles}). Here the Fermi gas is a normal fluid, and 
the dimensionless temperature $\theta(\mathbf{r})= k_B T/ k_F(\mathbf{r})$ is large, 
so that the Stefan-Boltzmann formulas are not applicable. Consequently, an 
asymptotic formula like \eqref{Eq:local_asymptotic_formula} cannot be derived 
for $S$ and $E-E_0$ of the whole trap.

\begin{figure}
\includegraphics[width=0.44\textwidth,clip]{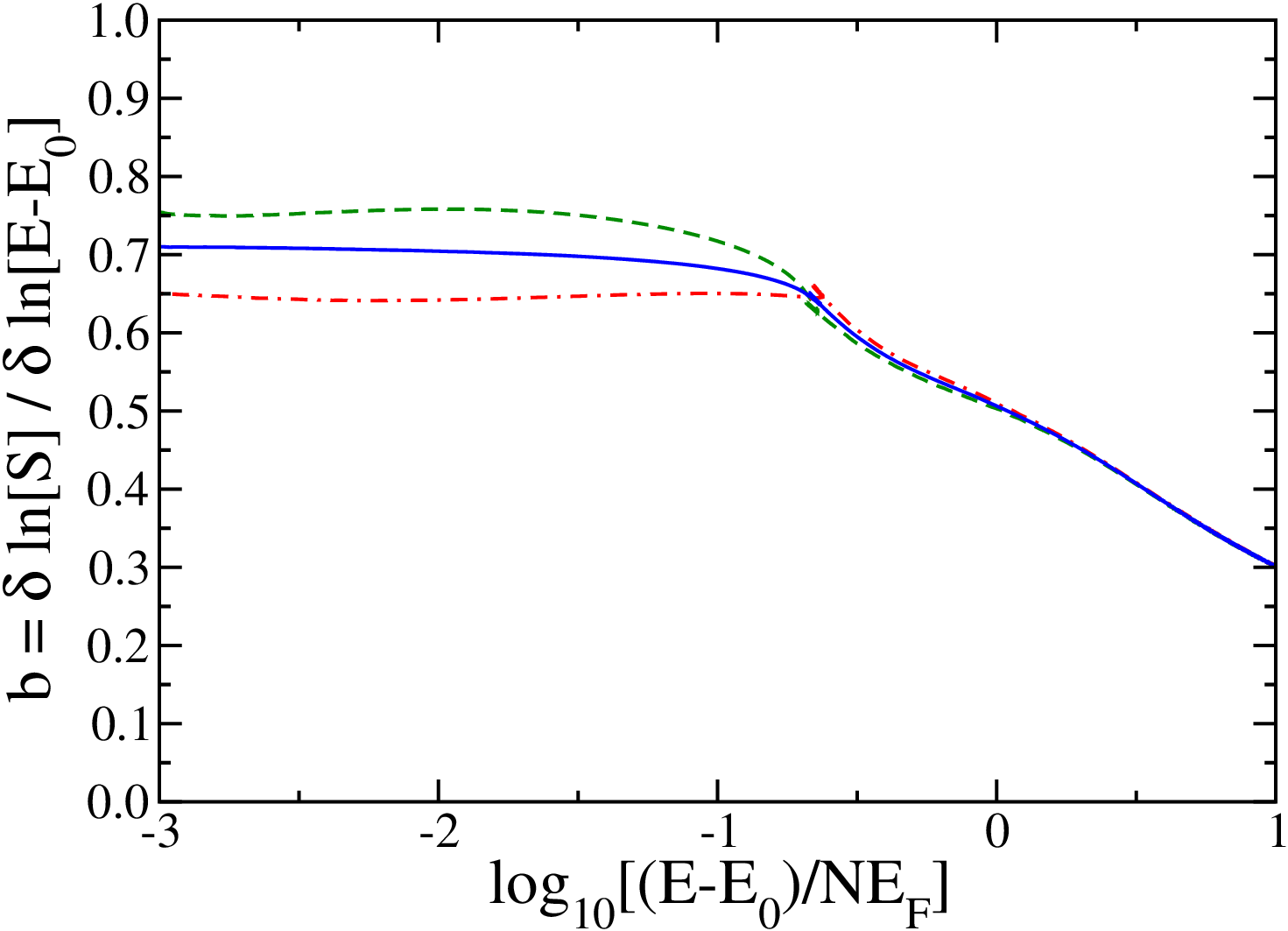}
\caption{(Color online) The local exponent $b=\partial \ln S/\partial \ln (E-E_0)$ 
of the function $S(E)$ where $E$ is two times the potential energy (green dashed), 
two times the internal energy (red dot-dashed), and one times the total energy 
(blue solid). Here $E_0$ is the ground-state energy for $T=0$ and $S=0$.}
\label{Fig:exponents}
\end{figure}

Empirically, it has been found in the experiments by Luo 
\textit{et al.}\ \cite{Luo07} that the total entropy at low energies varies with an 
effective power law
\begin{equation}
S/N k_B \sim [(E-E_0)/N E_F]^b 
\label{Eq:asymptotic_formula}
\end{equation}
with an exponent $b\approx 0.59$. In order to check whether 
such a behavior is consistent with a microscopic theory, we 
determine the exponent $b=\partial\ln S / \partial\ln (E-E_0)$ 
by logarithmic differentiation of the blue-green-red solid curve $S=S(E)$ in 
Fig.\ \ref{Fig:entropy}. The result is shown as blue solid curve in 
Fig.\ \ref{Fig:exponents} and confirms that $S(E)$ in a trap indeed exhibits 
a power law behavior in the regime near the ground state. 
We have adjusted the ground-state energy $E_0$ by fine 
tuning in order to obtain a well defined exponent in the limit $E\rightarrow E_0$. 
The resulting exponent $b=0.70$ is surprisingly close to the value $0.75$ of the 
local asymptotic formula \eqref{Eq:local_asymptotic_formula} but differs from 
the value $b=0.59\pm 0.03$ inferred from the experimental fit.

In Fig.\ \ref{Fig:exponents} the superfluid transition is located at the position
$\log_{10}[(E-E_0)/N E_F]=-0.65$. The left part of the blue solid curve corresponds 
to the superfluid region. Here the exponent is nearly constant up to the superfluid 
transition. Even though the Stefan-Boltzmann formulas are not valid, the exponent 
$b=0.70$ is remarkably close to the theoretical value $0.75$. The right part 
of the blue solid curve corresponds to the normal fluid region. Here the logarithmic 
derivative of the entropy with respect to energy decreases monotonically with 
increasing energy $E$ and thus no well defined exponent can be attributed to the 
high energy part of the curve. 

The virial theorem implies the energy relations $E=2 E_\mathrm{pot}=2 U$. In order
to check the validity of the virial theorem we consider the entropy functions
$S=S(E_\mathrm{pot})$ and $S=S(U)$ and calculate the related exponents 
$b(E_\mathrm{pot})= \partial\ln S / \partial\ln (E_\mathrm{pot} - E_{\mathrm{pot},0})$ 
and $b(U)= \partial\ln S / \partial\ln (U-U_0)$, which are shown in Fig.\ 
\ref{Fig:exponents} as green dashed line and as red dot-dashed line, respectively. 
These lines should be compared with the blue solid line, which represents the 
exponent $b(E)= \partial\ln S / \partial\ln (E-E_0)$. In the normal-fluid region 
the virial theorem is well satisfied, because the right parts of the curves are 
lying nearly on top of each other where the small deviations are numerical errors. 

In the superfluid region the virial theorem is satisfied only approximately because 
of the modification of the theory in order to have a gapless Bogoliubov-Anderson mode 
(see Sec.\ II.J in Ref.\ \onlinecite{Haussmann3}). As a consequence, the left parts 
of the curves deviate from each other. We find three different values for the, 
exponents which are $b(E_\mathrm{pot})=0.75$, $b(U)=0.65$, and 
$b(E)=0.70$. These results are related to the three different ground-state 
energies \eqref{Eq:potential_energy_3}, \eqref{Eq:internal_energy_3}, and 
\eqref{Eq:total_energy_3}, respectively.

\section{Conclusions}
\label{section_7}
Based on our previous results for the BCS to BEC crossover problem in a 
homogeneous gas \cite{Haussmann3}, we have calculated density and 
entropy profiles in a trap within a local density approximation.
In addition, we have determined the total entropy $S$ and energy $E$ in the 
unitary regime and have compared our results with both experiment and recent 
theories. For temperatures above the superfluid transition temperature,
very good agreement is obtained. However, the value of the critical 
temperature and the behavior at very low temperatures differ appreciably 
from the experimental estimates and their theoretical analysis in earlier work. 

First of all, our value for the Bertsch parameter $\xi$ and thus the 
ground-state energy $E_0$ is about $10\%$ smaller 
than the results obtained from variational Monte Carlo
calculations or from Gaussian fluctuation theories around the BCS
ansatz for the ground state. While our value $\xi=0.36$ agrees well 
with the most precise results so far obtained from an $\epsilon=4-d$ 
expansion \cite{Arnold07}, it differs from those obtained from variational 
Monte Carlo calculations \cite{Carlson03,Astra04}, or from those including 
Gaussian fluctuations around the BCS mean field \cite{HLD06,DSR08}, where 
$\xi=0.42(1)$ or $\xi=0.40$, respectively. Given the uncertainty in determining 
$\xi$ experimentally (which requires an extrapolation to zero temperature) 
it is clearly important  for theory to make precise predictions for $\xi$ which 
do not rely on approximations that are apparently limiting all present results. 
In view of the fundamental importance of this parameter in the context of 
strongly interacting Fermi gases, progress here would be highly desirable.

As a second point, our values for both the critical temperature and value of entropy 
at $T_c$ are appreciably lower than those obtained in the theories of Bulgac 
\textit{et al.}\ \cite{Bulgac07} and of Hu \textit{et al.}\ \cite{HDL07} and also 
those inferred from the original analysis of the experimental data \cite{Luo07}. 
Now, as is evident from Fig.\ \ref{Fig:entropy}, a measurement of the function 
$S(E)$ does not provide a  sensitive measure of the critical temperature. 
Quantitative results for the critical temperature of the unitary gas have been 
obtained by Shin \textit{et al.}\ \cite{Shin08}. They rely on using gases with 
a finite imbalance $n_{\uparrow}\ne n_{\downarrow}$, which have always a 
fully polarized outer shell in a trap. Since a single species ultracold
Fermi gas is noninteracting, its temperature
can be reliably determined from cloud profiles. An extrapolation back to
zero imbalance gives a critical temperature at unitarity which is close to the 
value predicted both in our theory and in Monte-Carlo calculations by Burovski 
\textit{et al.}\ \cite{Burovski06,Burovski08} and Bulgac \textit{et al.}\ 
\cite{Bulgac08}. As pointed out in section IV, there is now evidence that
our result $k_B T_c \simeq 0.21\, E_F$ for the critical temperature of the 
unitary gas in a trap is rather precise. Together with the entropy-temperature
curve shown in Fig.\ \ref{Fig:entropy-temperature}, this would allow doing
precise thermometry for balanced gases, that has not been possible so far. 

Finally, we have shown that for temperatures of order 
$k_B T \approx 0.06\, \varepsilon_F(\mathbf{0}) \approx 0.10\, E_F$
which have been reached experimentally \cite{Shin08}, an efficient way
of further cooling the gas is possible by removing the high entropy  outer part
of the atomic cloud and readjusting the confining potential. This method is similar
in spirit than the standard evaporative cooling idea, but potentially much more 
efficient. It might open the avenue to reach regimes in which the  entropy per 
particle is much less than $k_B$, a necessary condition for realizing many of the 
nontrivially ordered states that are in principle accessible with ultracold fermions
\cite{BDZ08}.

\acknowledgments
\noindent
We would like to thank A.\ Bulgac, P.D.\ Drummond, J.E.\ Drut and J.\ Thomas for 
useful discussions and for making their results available for comparison with our 
theory. W.Z.\ is grateful for the hospitality as a visitor at the MIT-Harvard 
Center for Ultracold Atoms and for a number of helpful discussions with M.\ Zwierlein.
This work was supported by the DFG Forschergruppe 801 on ultracold quantum gases.

\end{document}